

\def\singlespace{\normalbaselines}
\def\oneandahalfspace{\baselineskip=1.15\normalbaselineskip plus 1pt
\lineskip=2pt\lineskiplimit=1pt}

\def\np{\vfill\eject}
\def\nl{\hfil\break}

\def\nofirstpagenoten{\nopagenumbers\footline={\ifnum\pageno>1\tenrm
\hss\folio\hss\fi}}
\def\nofirstpagenotwelve{\nopagenumbers\footline={\ifnum\pageno>1\twelverm
\hss\folio\hss\fi}}
\def\leaderfill{\leaders\hbox to 1em{\hss.\hss}\hfill}
\def\ft#1#2{{\textstyle{{#1}\over{#2}}}}
\def\frac#1/#2{\leavevmode\kern.1em
\raise.5ex\hbox{\the\scriptfont0 #1}\kern-.1em/\kern-.15em
\lower.25ex\hbox{\the\scriptfont0 #2}}
\def\sfrac#1/#2{\leavevmode\kern.1em
\raise.5ex\hbox{\the\scriptscriptfont0 #1}\kern-.1em/\kern-.15em
\lower.25ex\hbox{\the\scriptscriptfont0 #2}}


\parindent=20pt
\def\narrow{\advance\leftskip by 40pt \advance\rightskip by 40pt}

\def\AB{\bigskip
        \centerline{\bf ABSTRACT}\medskip\narrow}
\def\nonarrower{\advance\leftskip by -40pt\advance\rightskip by -40pt}
\def\AE{\bigskip\nonarrower}

\def\boxit#1{\vbox{\hrule\hbox{\vrule\kern3pt
        \vbox{\kern3pt#1\kern3pt}\kern3pt\vrule}\hrule}}

\def\gtorder{\mathrel{\raise.3ex\hbox{$>$}\mkern-14mu
             \lower0.6ex\hbox{$\sim$}}}
\def\ltorder{\mathrel{\raise.3ex\hbox{$<$}|mkern-14mu
             \lower0.6ex\hbox{\sim$}}}
\def\dalemb#1#2{{\vbox{\hrule height .#2pt
        \hbox{\vrule width.#2pt height#1pt \kern#1pt
                \vrule width.#2pt}
        \hrule height.#2pt}}}

\font\fourteentt=cmtt10 scaled \magstep2
\font\fourteenbf=cmbx12 scaled \magstep1
\font\fourteenrm=cmr12 scaled \magstep1
\font\fourteeni=cmmi12 scaled \magstep1
\font\fourteenss=cmss12 scaled \magstep1
\font\fourteensy=cmsy10 scaled \magstep2
\font\fourteensl=cmsl12 scaled \magstep1
\font\fourteenex=cmex10 scaled \magstep2
\font\fourteenit=cmti12 scaled \magstep1
\font\twelvett=cmtt10 scaled \magstep1 \font\twelvebf=cmbx12
\font\twelverm=cmr12 \font\twelvei=cmmi12
\font\twelvess=cmss12 \font\twelvesy=cmsy10 scaled \magstep1
\font\twelvesl=cmsl12 \font\twelveex=cmex10 scaled \magstep1
\font\twelveit=cmti12
\font\tenss=cmss10
 
 \font\ninebf=cmbx7 scaled \magstep1
\font\ninerm=cmr7 scaled \magstep1 \font\ninei=cmmi7 scaled \magstep1
\font\ninesy=cmsy7 scaled \magstep1 
\font\eightrm=cmr7 scaled 1140 
 
\font\sevenbf=cmbx7 \font\sevenrm=cmr7 \font\seveni=cmmi7
\font\sevensy=cmsy7 

\catcode`@=11
\newskip\ttglue
\newfam\ssfam

\def\fourteenpoint{\def\rm{\fam0\fourteenrm}
\textfont0=\fourteenrm \scriptfont0=\tenrm \scriptscriptfont0=\sevenrm
\textfont1=\fourteeni \scriptfont1=\teni \scriptscriptfont1=\seveni
\textfont2=\fourteensy \scriptfont2=\tensy \scriptscriptfont2=\sevensy
\textfont3=\fourteenex \scriptfont3=\fourteenex \scriptscriptfont3=\fourteenex
\def\it{\fam\itfam\fourteenit} \textfont\itfam=\fourteenit
\def\sl{\fam\slfam\fourteensl} \textfont\slfam=\fourteensl
\def\bf{\fam\bffam\fourteenbf} \textfont\bffam=\fourteenbf
\scriptfont\bffam=\tenbf \scriptscriptfont\bffam=\sevenbf
\def\tt{\fam\ttfam\fourteentt} \textfont\ttfam=\fourteentt
\def\ss{\fam\ssfam\fourteenss} \textfont\ssfam=\fourteenss
\tt \ttglue=.5em plus .25em minus .15em
\normalbaselineskip=16pt
\abovedisplayskip=16pt plus 4pt minus 12pt
\belowdisplayskip=16pt plus 4pt minus 12pt
\abovedisplayshortskip=0pt plus 4pt
\belowdisplayshortskip=9pt plus 4pt minus 6pt
\parskip=5pt plus 1.5pt
\setbox\strutbox=\hbox{\vrule height12pt depth5pt width0pt}
\let\sc=\tenrm
\let\big=\fourteenbig \normalbaselines\rm}
\def\fourteenbig#1{{\hbox{$\left#1\vbox to12pt{}\right.\n@space$}}}

\def\twelvepoint{\def\rm{\fam0\twelverm}
\textfont0=\twelverm \scriptfont0=\ninerm \scriptscriptfont0=\sevenrm
\textfont1=\twelvei \scriptfont1=\ninei \scriptscriptfont1=\seveni
\textfont2=\twelvesy \scriptfont2=\ninesy \scriptscriptfont2=\sevensy
\textfont3=\twelveex \scriptfont3=\twelveex \scriptscriptfont3=\twelveex
\def\it{\fam\itfam\twelveit} \textfont\itfam=\twelveit
\def\sl{\fam\slfam\twelvesl} \textfont\slfam=\twelvesl
\def\bf{\fam\bffam\twelvebf} \textfont\bffam=\twelvebf
\scriptfont\bffam=\ninebf \scriptscriptfont\bffam=\sevenbf
\def\tt{\fam\ttfam\twelvett} \textfont\ttfam=\twelvett
\def\ss{\fam\ssfam\twelvess} \textfont\ssfam=\twelvess
\tt \ttglue=.5em plus .25em minus .15em
\normalbaselineskip=14pt
\abovedisplayskip=14pt plus 3pt minus 10pt
\belowdisplayskip=14pt plus 3pt minus 10pt
\abovedisplayshortskip=0pt plus 3pt
\belowdisplayshortskip=8pt plus 3pt minus 5pt
\parskip=3pt plus 1.5pt
\setbox\strutbox=\hbox{\vrule height10pt depth4pt width0pt}
\let\sc=\ninerm
\let\big=\twelvebig \normalbaselines\rm}
\def\twelvebig#1{{\hbox{$\left#1\vbox to10pt{}\right.\n@space$}}}

\def\tenpoint{\def\rm{\fam0\tenrm}
\textfont0=\tenrm \scriptfont0=\sevenrm \scriptscriptfont0=\fiverm
\textfont1=\teni \scriptfont1=\seveni \scriptscriptfont1=\fivei
\textfont2=\tensy \scriptfont2=\sevensy \scriptscriptfont2=\fivesy
\textfont3=\tenex \scriptfont3=\tenex \scriptscriptfont3=\tenex
\def\it{\fam\itfam\tenit} \textfont\itfam=\tenit
\def\sl{\fam\slfam\tensl} \textfont\slfam=\tensl
\def\bf{\fam\bffam\tenbf} \textfont\bffam=\tenbf
\scriptfont\bffam=\sevenbf \scriptscriptfont\bffam=\fivebf
\def\tt{\fam\ttfam\tentt} \textfont\ttfam=\tentt
\def\ss{\fam\ssfam\tenss} \textfont\ssfam=\tenss
\tt \ttglue=.5em plus .25em minus .15em
\normalbaselineskip=12pt
\abovedisplayskip=12pt plus 3pt minus 9pt
\belowdisplayskip=12pt plus 3pt minus 9pt
\abovedisplayshortskip=0pt plus 3pt
\belowdisplayshortskip=7pt plus 3pt minus 4pt
\parskip=0.0pt plus 1.0pt
\setbox\strutbox=\hbox{\vrule height8.5pt depth3.5pt width0pt}
\let\sc=\eightrm
\let\big=\tenbig \normalbaselines\rm}
\def\tenbig#1{{\hbox{$\left#1\vbox to8.5pt{}\right.\n@space$}}}
\let\rawfootnote=\footnote \def\footnote#1#2{{\rm\parskip=0pt\rawfootnote{#1}
{#2\hfill\vrule height 0pt depth 6pt width 0pt}}}

\def\tenfoot{\tenpoint\hskip-\parindent\hskip-.1cm}

\twelvepoint
\def\sbullet{\raise.2em\hbox{$\scriptscriptstyle\bullet$}}
\nofirstpagenotwelve
\hsize=16.5 truecm
\baselineskip 15pt

\def\ft#1#2{{\textstyle{{#1}\over{#2}}}}

\def\lagr{{\cal L}}
\def\tJ{\widetilde J}
\def\th{\tilde h}
\def\tB{\widetilde B}

\oneandahalfspace
\rightline{CTP TAMU--82/91}
\rightline{October 1991}

\vskip 2truecm
\centerline{\bf Anomaly-free $W$-gravity Theories}
\vskip 1.5truecm
\centerline{C.N. Pope\footnote{$^\star$}{\tenfoot Supported in part by the
U.S. Department of Energy, under
grant DE-FG05-91ER40633.}\footnote{}{\tenfoot Contribution to the
proceedings of {\it Strings and Symmetries}, Stony Brook, 1991.}}

\vskip 1.5truecm

\centerline{\it Center for Theoretical Physics, Texas A\&M University,}
\centerline{\it College Station, TX 77843--4242, USA.}

\vskip 1.5truecm
\AB\singlespace
      \AE\oneandahalfspace
     We give a review of some recent developments in the quantisation of
$W$-gravity theories.  In particular, we discuss the construction of
anomaly-free $W_\infty$ and $W_3$ gravities.

\np
\noindent
{\bf 1. Introduction}
\bigskip

   Two-dimensional gravity has provided a rich and fascinating field of study
in the last few years.  One can view two-dimensional gravity as being the gauge
theory of the Virasoro algebra in two dimensions.  Since higher-spin
extensions of the Virasoro algebra exist, it is natural to investigate the
corresponding two-dimensional gauge theories.  Many of these higher-spin
algebras, known as $W$ algebras, have been discovered.  The first, called
$W_3$, contains a current of spin 3 in addition to the usual spin-2 current,
the holomorphic energy-momentum tensor, of the Virasoro algebra [1].
Subsequently, generalisations to $W_N$ algebras were obtained, which contain
currents of each spin $s$ in the interval $2\le s \le N$ [2].  A general
feature of these extended algebras is that they are non-linear, in the sense
that the operator product of two currents may produce terms that must be
viewed as composite operators, built from products of the fundamental
currents in the algebra.  The reason for this is that to leading order, the
OPE of currents with spins $s$ and $s'$ gives a quantity of spin
$s''=s+s'-2$, which may exceed $N$ if $N>2$.  This argument for the
occurrence of non-linearities breaks down if $N=\infty$, and indeed a linear
algebra of this kind, known as $W_\infty$, exists [3].  In most other
respects, $W_\infty$ is qualitatively similar to the finite-$N$ algebras.
For example, they all have non-trivial central terms in the OPE of any pair
of equal-spin currents. There is another related algebra, known as
$W_{1+\infty}$, which has currents of each spin $1\le s\le \infty$ [3,4].
All of the $W_N$ algebras, with $N$ finite or infinite, admit contractions
to algebras that may, in a sense to be clarified below, be thought of as
classical limits.  In the case of $W_\infty$, the corresponding contracted
algebra is the $w_\infty$ algebra found in [5]. For $W_{1+\infty}$, its
contraction, $w_{1+\infty}$, is isomorphic to the algebra of area-preserving
diffeomorphisms on a cylinder [3,5].

     In this article, we shall review the construction and quantisation of two
$W$-gravity theories.  The first of these is based upon the classical
two-dimensional gauge theory of the $w_\infty$ algebra [6].  We shall see that
in order to quantise the theory, it is necessary to renormalise the classical
currents that generate the classical $w_\infty$ symmetry.  After the dust has
settled, one finds that the renormalised currents generate the full $W_\infty$
algebra [7].  Thus the classical $w_\infty$ gravity that one starts with ends
up
as quantum $W_\infty$ gravity.  The second example that we shall consider here
is that of $W_3$ gravity.  One might think that the non-linearities of the
$W_3$ algebra would make the analysis of this case much more difficult.  Up to
a point, this is true.  However, the essential qualitative features are in fact
remarkably similar to those of the $W_\infty$ case.  The starting point in this
second example is the two-dimensional gauge theory of a contraction of the
$W_3$ algebra [8,9].  It would therefore be more appropriate to call this
classical $w_3$ gravity, rather than $W_3$ gravity.  Again, we shall see that
the quantisation process requires that the classical $w_3$ currents must be
renormalised.  In the end, the renormalised currents generate the full $W_3$
algebra [10,11].

\vfill\eject

\bigskip
\noindent{\bf 2. Classical and Quantum $W_\infty$ Gravity}
\bigskip

     The classical theory of $w_\infty$ gravity was constructed in [6].  In its
simplest form, one can consider a {\it chiral} gauging of $w_\infty$; {\it
i.e.}\ one gauges just one copy of the algebra, in, say, the left-moving sector
of the two-dimensional theory.  We shall discuss non-chiral gaugings in more
detail later.  For now we just remark that, thanks to an ingenious trick
introduced in [9], involving the use of auxiliary fields, the treatment of the
non-chiral case can be essentially reduced to two independent copies of the
chiral case.

     As our starting point, let us consider the free action $S=1/\pi \int
d^2z L$ for a single scalar field in two dimensions, where $L$ is given by
$$
L=\ft12 \bar\partial \varphi\, \partial\varphi. \eqno(2.1)
$$
Here, we use coordinates $z=x^-$ and $\bar z=x^+$ on the (Euclidean-signature)
worldsheet. This action is invariant under the semi-rigid spin-$s$
transformations
$$
\delta\varphi=\sum_{s\ge2} k_s (\partial\varphi)^{s-1},\eqno(2.2)
$$
where the parameters $k_s$ are functions of $z$, but not $\bar z$.  These
transformations are generated by the currents
$$
v_s(z)={1\over s} (\partial\varphi)^s.\eqno(2.3)
$$
At the classical level, these currents generate the $w_\infty$ algebra.  In
operator-product language, this means that they close on the $w_\infty$ algebra
at the level of {\it single} contractions.  To keep track of the orders it is
useful to introduce Planck's constant, so that the OPE of the field $\varphi$
is
$$
\partial\varphi(z)\partial\varphi(w)\sim {\hbar\over(z-w)^2}.\eqno(2.4)
$$
The OPEs of the currents are then given by
$$
\hbar^{-1} v_s(z) v_{s'}(w)\sim (s+s'-2){v_{s+s'-2}(w)\over (z-w)^2}+ (s-1)
{\partial v_{s+s'-2}(w)\over z-w} +O(\hbar).\eqno(2.5)
$$
The $\hbar$-independent terms on the right-hand side are precisely those for
the $w_\infty$ algebra.

     The classical semi-rigid $w_\infty$ symmetry (2.2) of (2.1) can be gauged
by introducing a gauge field $A_s$ for each current $v_s$.  Thus we find that
the Lagrangian
$$
L=
\ft12 \bar\partial \varphi\, \partial\varphi-\sum_{s\ge2} A_s v_s\eqno(2.6)
$$
is invariant under local $w_\infty$ transformations [6], where the gauge fields
are assigned the transformation rules:
$$
\delta A_s=\bar\partial k_s -\sum_{s'=2}^s\Big( (s'-1)A_{s'} \partial
k_{s-s'+2}
-(s-s'+1)k_{s-s'+2}\partial A_{s'}\Big).\eqno(2.7)
$$

     When one is presented with a classical theory it is natural, when
considering its quantisation, to begin by contemplating what might go wrong.
In the case of a gauge theory, with classical local symmetries, the
obvious danger is that these might become anomalous at the quantum level.
Indeed, in the case of two-dimensional gravity, the gauge theory of the
Virasoro algebra, we know that anomaly freedom requires that the matter fields
should generate the Virasoro algebra with central charge $c=26$, in order to
cancel the central-charge contribution of $-26$ from the ghosts for the gauge
fixing of the spin-2 gauge field (the metric).  We can certainly expect to meet
analogous anomalies in the higher-spin generalisations that we are considering
here.  In fact, potentially-worse things could also happen:

     Commonly, as for
example in the case of the Virasoro algebra and two-dimensional gravity, one
has matter currents that are quadratic in matter fields.  These, by
construction, generate the gauge algebra at the classical (single-contraction)
level.  Upon quantisation, higher numbers of contractions must also be taken
into account, corresponding to Feynman diagrams with closed loops.  If the
currents are at most quadratic in matter fields, then the ``worst case'' is to
have two contractions between a pair of currents.  This corresponds therefore
to
a pure $c$-number term in the OPE of currents; in fact, the central term in the
Virasoro algebra.  It is these terms that in fact save the critical
two-dimensional gravity theory from anomalies, by cancelling against anomalies
from the ghost sector.

     Things are potentially worse in the case of $w_\infty$ gravity because now
the currents (2.3) involve arbitrarily-high powers of the matter field
$\varphi$.  Thus at the quantum level, one might get matter-dependent
anomalies, associated with diagrams corresponding to multiple contractions of
the currents (2.3) that still have some matter fields left uncontracted.  Of
course the question of whether a theory is actually anomalous is really a
cohomological one, in the sense that the crucial question is whether or not it
is possible to introduce finite local counterterms, and $\hbar$-dependent
renormalisations of the classical transformation rules in such a way as to
remove the apparently-anomalous contributions of the kind we have been
considering.  Only if such an attempt fails can the theory be said to be
anomalous.

     In general, the process of quantising a theory, and introducing
counterterms and renormalisations of the transformation rules order by order in
$\hbar$ to remove potential anomalies, can be a complicated and tedious one.
Fortunately, in our two-dimensional example the process of removing potential
matter-dependent anomalies can be accomplished in one fell swoop.  All that we
have to do is to find quantum renormalisations of the classical currents (2.3)
such that at the full quantum level (arbitrary numbers of contractions in the
OPEs) they close on an algebra.  This algebra, whatever it turns out to be,
will be the quantum renormalisation of the original classical $w_\infty$
algebra.  In fact, as we shall see, it is precisely $W_\infty$ [7].

     The problem, then, boils down to the fact that the currents (2.3) do not
close on any algebra at the quantum level.  We must therefore seek
$\hbar$-dependent modifications of them, with a view to achieving quantum
closure.  The most general plausible modifications would consist of
higher-order terms added to (2.3) in which the same number of derivatives (to
give the same spin $s$) are distributed over smaller numbers of $\varphi$
fields.  From (2.4), we see that $\varphi$ has the dimensions of $\sqrt\hbar$,
and so the modifications will be of the form of power series in $\sqrt\hbar$.
Thus we may try an ansatz of the form
$$
V_s={1\over s} (\partial\varphi)^s + \alpha_s \sqrt\hbar
(\partial\varphi)^{s-2} \partial^2\varphi + \beta_s \hbar (\partial
\varphi)^{s-3} \partial^3\varphi +\gamma_s \hbar (\partial\varphi)^{s-4}
(\partial^2\varphi)^2 + O(\hbar^{3/2}),\eqno(2.8)
$$
for constant coefficients $\alpha_s$, $\beta_s$, $\gamma_s,\cdots$ to be
determined.  Requiring quantum closure of the OPE algebra for these currents
will then give conditions on this infinite number of coefficients.  They
will not be determinable uniquely, since one is always free to make
redefinitions of currents of the form $V_s\to V_s +\partial V_{s-1}+\cdots$.
However, if, for convenience and without loss of generality, we demand that the
currents should all be quasiprimary with respect to the energy-momentum tensor
$V_2$, then the result is unique.  The expressions for the first few
renormalised currents (spins 2, 3 and 4) are [7]:
$$
\eqalign{
V_2&=\ft12 (\partial\varphi)^2 +\ft12\sqrt\hbar \partial^2\varphi,\cr
V_3&=\ft13(\partial\varphi)^3+\ft12 \sqrt\hbar \partial\varphi
\partial^2\varphi
+\ft1{12}\hbar \partial^3\varphi,\cr
V_4&=\ft14(\partial\varphi)^4 +\ft12\sqrt\hbar
(\partial\varphi)^2\partial^2\varphi -\ft1{20}\hbar (\partial^2\varphi)^2
+\ft15 \hbar \partial\varphi \partial^3\varphi +\ft1{60}\hbar^{3/2}
\partial^4\varphi.\cr}\eqno(2.9)
$$

     These renormalised currents can be recognised as the currents of the
$W_\infty$ algebra in the following way:  We know that there is a realisation
of $W_{1+\infty}$ in terms of bilinear currents built from a free complex
fermion [12].  We also know that one can make a one-parameter family of
rotations of the basis for $W_{1+\infty}$, retaining the quasiprimary nature
of all the currents, such that for one member of the family, the spin-1
current can be truncated out, leaving currents of spins $\ge2$ that generate
$W_\infty$ [13,14].  The first few such currents are:
$$
\eqalign{
V_2&=\partial\bar\psi\, \psi,\cr
V_3&=\ft12 \partial^2\bar\psi\, \psi -\ft12\partial\bar\psi\, \partial\psi,\cr
V_4&=\ft15 \partial^3\bar\psi\,\psi-\ft35 \partial^2\bar\psi\, \partial\psi +
\ft15 \partial\bar\psi\, \partial^2\psi.\cr}\eqno(2.10)
$$
It is then straightforward to see that the currents (2.9) are nothing but the
bosonisation of the fermionic currents (2.10) [15,7], where we write
$$
\psi=:e^{\varphi/\sqrt\hbar}:,\qquad \qquad \bar\psi=:e^{-\varphi/\sqrt\hbar}:.
\eqno(2.11)
$$

     Having obtained renormalised currents that close on an algebra (the
$W_\infty$ algebra) at the full quantum level, we are now guaranteed to
have a quantum theory with no matter-dependent anomalies.  The prescription
for writing down the counterterms and renormalisations of the
transformation rules (2.2) and (2.7) necessary to make this anomaly freedom
manifest is very straightforward.  For the counterterms, we simply replace
the classical currents $v_s$ in the Lagrangian (2.6) by the renormalised
currents $V_s$, of which the first few are given by (2.9).  The
$\hbar$-dependent terms are the necessary counterterms.  For the
transformation rules, we use the ones that are generated by the
renormalised currents.  For $\varphi$, this means we have
$$
\delta\varphi=\hbar^{-1}\sum_{s\ge2} \oint{dz\over 2\pi i} k_s(z) V_s(z)
\varphi(w).\eqno(2.12)
$$
For $A_s$, we will have
$$
\delta A_s=\bar\partial k_s +\hat\delta A_s,\eqno(2.13)
$$
where $\hat\delta A_s$ is such that
$$
\sum_{s\ge 2}\int\Big( \hat\delta A_s V_s +A_s\delta V_s\Big),\eqno(2.14)
$$
with $\delta V_s$ given by
$$
\delta V_s=\hbar^{-1} \sum_{s'\ge 2}\oint{dz\over 2\pi i} k_{s'}(z)
V_{s'}(z) V_s(w).\eqno(2.15)
$$
The $\hbar$-independent terms in these transformation rules are precisely
the original classical ones (2.2) and (2.7).  The $\hbar$-dependent terms
are the renormalisations necessary, together with the counterterms, to make
the absence of matter-dependent anomalies manifest.  The fact that the
potential matter-dependent anomalies are actually removable by this means, {\it
i.e.}\ that they are cohomologically trivial, is a consequence the fact that it
is possible to renormalise the classical currents (2.3) to give currents that
{\it do} close at the quantum level.

     So far, we have been concerned here only with the question of
matter-dependent anomalies.  This is an issue that does not even arise for
usual formulations of two-dimensional gravity, since the Virasoro symmetry
is usually realised linearly.  We must still face the analogue of the
anomaly that one {\it does} meet in two-dimensional gravity, namely the
universal anomaly that is cancelled by choosing a $c=26$ matter realisation
in order to cancel against the $-26$ contribution to the total central
charge coming from the gravity ghosts.  For $W_\infty$ gravity, we face a
more serious-looking problem, since now there will be ghosts associated
with the fixing of the gauge symmetry for each of the gauge-fields $A_s$.
The ghosts for a spin-$s$ gauge field contribute
$$
c_{\rm gh}(s)=-12s^2+12s-2\eqno(2.16)
$$
to the ghostly central charge.  Summing over all $s\ge2$ would seem to imply
that the total ghostly central charge is $c_{\rm gh}({\rm tot})=-\infty$.
At best, an infinity of matter fields seem to be needed, and even then, the
process of cancelling the universal anomaly could be a delicate one.  There
is, however, a different approach that one could take.  The sum of $c_{\rm
gh}(s)$ over all $s\ge 2$ can be regularised using zeta-function techniques,
to give
$$
\eqalign{
c_{\rm gh}({\rm tot})&=-\sum_{n\ge0}\Big(6(n+\ft32)^2-\ft12\Big) \cr
&=-6\zeta(-2,\ft32)+\ft12 \zeta(0,\ft32)\cr
&=2,\cr}\eqno(2.17)
$$
where $\zeta(s,a)$ is the generalised Riemann zeta function, defined by
analytic continuation from $\zeta(s,a)=\sum_{n\ge0}(n+a)^{-s}$ in the region
${\cal R}(s)>1$.  Of course this regularisation procedure looks somewhat
arbitrary, but actually there is very strong evidence to suggest that there
is an underlying justification for it.  The ghost anomalies for $W_\infty$
have been analysed in detail in a BRST approach in [16,17].  Not only does
one have an anomaly in the spin-2 sector, with central charge given by
(2.17), but also there are related anomalies in all the higher-spin sectors,
corresponding to the central terms in the OPEs of each higher-spin current
with itself.  The relative values of all these central terms are related in
the $W_\infty$ algebra, with just one overall scale parameter free (the
central charge $c$). Thus for the ghost currents to provide a realisation of
the algebra, it is necessary that all the higher-spin anomalies must
regularise in a self-consistent way.  This was examined up to the spin-18
level in [17], and it was found that the only plausible regularisation
scheme was the one that gives $c_{\rm gh}({\rm tot})=2$.  One may hope that
there is some (higher-dimensional?) explanation for this that will
eventually emerge.

     Assuming for now that the $c_{\rm gh}({\rm tot})=2$ result is to be taken
seriously for $W_\infty$, it follows that the cancellation of the universal
anomalies will occur provided that the matter realisation of $W_\infty$ has
central charge $c_{\rm mat}=-2$.  Remarkably, this is precisely what we have
for
our single-scalar realisation!  One can easily check from (2.9) that the
background-charge term is precisely such as to give $c=-2$.  Thus, in a
regularised sense at least, the $W_\infty$ gravity theory that we have
constructed is free of all anomalies [7].  This includes not only the spin-2
anomaly, of the form $(c_{\rm gh}({\rm tot})+c_{\rm mat})\int k_2\partial^3
A_2$, but also the higher-spin anomalies, of the form $C_s\int k_s
\partial^{s+1} A_s$.  For the same reasons as discussed above, all of the
coefficients $C_s$ will vanish simultaneously, provided that $c_{\rm mat}$
is equal to $-2$.

     For now, the possible cancellation of the regularised universal anomlies
should perhaps be viewed as an amusing observation that may ultimately turn out
to have some deeper underlying explanation.  Perhaps the more important lesson
to be derived from looking at the quantisation of classical $w_\infty$ gravity
is that the key requirements are that one should be able to renormalise the
classical currents so that they close on an algebra at the full quantum level.
This ensures the absence of matter-dependent anomalies.   Furthermore, if the
central charge for the matter currents is chosen to cancel that from the ghosts
for the gauge fields, then the universal anomalies will cancel also.  These
{\it desiderata} can be summarised succinctly in one equation:  we require that
the BRST operator $Q$ should be nilpotent.

\bigskip
\noindent{\bf 3. Classical and Quantum $W_3$ gravity}
\bigskip

    The philosphy for quantising classical $w_3$ gravity is essentially the
same as that of the previous section.  The starting point is the classical
matter Lagrangian [8]
$$
L=\ft12\bar\partial \varphi_i \,\partial\varphi_i -h T- B W,\eqno(3.1)
$$
where $\varphi_i$ denotes a set of $n$ real matter fields; $h$ and $B$ are
spin-2 and spin-3 gauge fields; and the spin-2 and spin-3 matter currents $T$
and $W$ are given by
$$
\eqalign{
T&=\ft12 \partial\varphi_i \,\partial\varphi_i,\cr
W&=\ft13 d_{ijk}\partial\varphi_i  \,\partial\varphi_j\,\partial\varphi_k.\cr}
\eqno(3.2)
$$
The quantity $d_{ijk}$ is a totally-symmetric constant tensor that satisfies
$$
d_{(ij}{}^k d_{\ell m)k}=\mu \delta_{(ij}\delta_{\ell m)}.\eqno(3.3)
$$
    At the classical level (single contractions), the currents generate what
we may call the $w_3$ algebra,
$$
\eqalign{
\hbar^{-1}T(z)T(w)&\sim {\partial T\over z-w} +{2T\over (z-w)^2} +O(\hbar),\cr
\hbar^{-1}T(z)W(w)&\sim {\partial W\over z-w} +{3W\over (z-w)^2} +O(\hbar),\cr
\hbar^{-1}W(z)W(w)&\sim {\partial \Lambda \over z-w} + {2\Lambda \over
(z-w)^2}, \cr}\eqno(3.4)
$$
where $\Lambda$ is the composite current
$$
\Lambda=2\mu (T T).\eqno(3.5)
$$
This $w_3$ algebra is a classical limit of the full $W_3$ algebra, which is
given below.

     Various possible solutions for the tensor $d_{ijk}$, satisfying (3.3),
have been found [18].  They fall into two categories.  The first consists of
solutions for an arbitrary number of scalars $n$, with the components of the
(totally symmetric) tensor $d_{ijk}$ given by
$$
d_{111}=n,\qquad\qquad d_{1ab} =-n \delta_{ab};\quad (a=2,\ldots,n).\eqno(3.6)
$$
This satisfies (3.3) with $\mu=n^2$.  The second category of solution relies
upon the abnormalities and perversities of the Jordan algebras.  There are four
such solutions, with $n=5$, $8$, $14$ and $26$ scalars, corresponding to
invariant tensors of Jordan algebras over the reals, complex numbers,
quaternions and octonions respectively [18].  For the complex case, with $n=8$,
the $d_{ijk}$ tensor coincides with the symmetric $d_{ijk}$ tensor of  $SU(3)$.

    As in the case of the currents (2.3) that generate the $w_\infty$
contraction of $W_\infty$ classically, so also here the currents (3.2)
generate the $w_3$ contraction of $W_3$ classically.  At the full quantum level
of multiple contractions in the operator-product expansion, one finds that the
classical $w_3$ currents (3.2) fail to close on any algebra.  This is the
signal
for potential trouble with matter-dependent anomalies upon quantisation of the
theory.  The remedy is again to look for quantum renormalisations of the
currents (3.2) to give currents that do generate an algebra at the quantum
level.  Modulo the freedom to make field redefinitions, the answer, as in the
$w_\infty$ case, is unique.  In this case, it turns out that the resulting
algebra on which the renormalised currents close is $W_3$.  This algebra takes
the form [1] $$
\eqalignno{
\hbar^{-1}T(z) T(w)& \sim {\partial T(w)
\over z-w} +{2 T(w)\over (z-w)^2} +\hbar{c/2\over
(z-w)^4}&(3.7a) \cr
\hbar^{-1}T(z) W(w)&\sim {\partial W
\over z-w} +{3 W(w)\over (z-w)^2}&(3.7b)\cr
\hbar^{-1}W(z)W(w)& \sim {1\over z-w}\left(
\ft1{15}\hbar\partial^3 T(w) +{16\over 22+5c}
\partial \Lambda(w)\right) \cr
&+{1\over (z-w)^2} \left(
\ft3{10}\hbar\partial^2 T(w) +{32\over 22+5c}
 \Lambda(w)\right)\cr
&+\hbar{\partial T(w)\over (z-w)^3} +\hbar {2 T(w)\over
(z-w)^4} +\hbar^2{c/3 \over(z-w)^6}&(3.7c)\cr}
$$
In (3.7$c$), $\Lambda$ is a composite current;
$$
\Lambda=(TT) -\ft3{10}
\hbar \partial^2 T,\eqno(3.8)
$$
where the normal ordering is taken with respect to the modes of the currents
$T$, according to the prescription [19]
$$
(AB)(w)\equiv \oint {dz \over z-w}A(z)B(w).\eqno(3.9)
$$

    The possible renormalisations of the currents (3.2) can be parametrised by
$$
\eqalign{
T&=\ft12\partial\varphi^i \partial\varphi^i +\sqrt\hbar \alpha_i
\partial^2\varphi^i,\cr
W&=\ft13 d_{ijk}\partial\varphi^i \partial\varphi^j \partial
\varphi^k +\sqrt\hbar e_{ij} \partial\varphi^i \partial^2\varphi^j +\hbar f_i
\partial^3 \varphi^i,\cr}\eqno(3.10)
$$
The requirement that the currents generate the $W_3$ algebra gives a set of
conditions on the coefficients $d_{ijk}$, $\alpha_i$, $e_{ij}$ and $f_i$ that
may be found in [18].  The upshot is that the general family of classical
currents, with $d_{ijk}$ given by (3.6) for the case of $n$ scalars, can be
successfully renormalised to give currents that close at the quantum level, on
the $W_3$ algebra [18].  We shall give the form of the renormalised currents
below.  For the four exceptional cases based on the Jordan algebras, however,
it
appears that it is not possible to renormalise the currents with
$\hbar$-dependent corrections so as to achieve closure [18,20].  This includes
the special 8-scalar realisation based on the totally-symmetric $d_{ijk}$
tensor
of $SU(3)$. There is thus no sense in which the currents in these exceptional
cases could be said to be $W_3$ currents. Consequently the quantum theory of
$w_3$ gravity for any of the four exceptional cases will definitely suffer from
anomalies.  On the other hand, for the general family of $n$-scalar
realisations
with $d_{ijk}$ given by (3.6), as we shall see below, all anomalies can be
cancelled.

      Having established that the matter currents for the $n$-scalar
realisations (3.2), (3.6) can be renormalised to give currents that close, on
the $W_3$ algebra, we are now able to proceed to the next stage in the
quantisation procedure, by looking for a nilpotent BRST charge.  Fortunately
the construction has already been carried out for $W_3$ [21].  Despite the
non-linearities of the algebra, it turns out that the structure of the BRST
charge is quite similar to that for a linear algebra.  The main difference is
that now the ghost currents involve matter currents.  Thus we still have [21]
$$
Q=\oint dz \Big(c(T+\ft12 T_{\rm gh})+\gamma(W+ \ft12
W_{\rm gh})\Big),\eqno(3.11)
$$
where $T$ and $W$ are the matter currents that generate the $W_3$ algebra;
$T_{\rm gh}$ and $W_{\rm gh}$ are the ghost currents; and the ghost-antighost
pairs ($c$,$b$) and ($\gamma$,$\beta$) correspond respectively to the $T$ and
$W$ generators.   For nilpotency, we can see from (2.16) that the central
charge for the matter currents must be
$$
c=26+74=100.\eqno(3.12)
$$
{}From now on, we shall assume therefore that the matter currents satisfy the
$W_3$ algebra (3.7$a$-$c$) with $c=100$.  The ghost currents $T_{\rm gh}$ and
$W_{\rm gh}$ are then given by [21]
$$
\eqalignno{
T_{\rm gh}&=-2b\,\partial c-\partial
b\, c-3\beta\, \partial\gamma-2\partial\beta\, \gamma&(3.13a)\cr
W_{\rm gh}&=-\partial\beta\,
c-3\beta\, \partial c-\ft8{261}\big[\partial(b\, \gamma\,  T )+b\,
\partial\gamma \, T\big]\cr &\ \ +\ft{25}{6\cdot261}\hbar
\Big(2\gamma\, \partial^3b+9\partial\gamma\, \partial^2b
+15\partial^2\gamma\,\partial b+10\partial^3\gamma\,  b\Big).&(3.13b)\cr}
$$
Note that the spin-3 ghost current $W_{\rm gh}$ involves the spin-2 matter
current $T$.  This looks intuitively reasonable; one can view the non-linear
terms on the right-hand side of (3.7$c$) as being like a linear
algebra but with $T$-dependent structure ``constants.''  These structure
constants then appear in the construction of the ghost currents.  Note also
that the ghost currents need not, and indeed do not, satisfy the $W_3$ algebra
[21,22].  It is shown in [21,22] that, provided the matter central charge is
given by (3.12), the BRST operator (3.11) is indeed nilpotent.

     The remaining ingredient needed for constructing anomaly-free $W_3$
gravity is a matter realisation of the $W_3$ algebra with central charge
$c=100$.  A 2-scalar realisation, with background charge that can be tuned to
give, in particular, $c=100$, was obtained in [2] by making use of the quantum
Miura transformation.  The most general known realisations in terms of scalar
fields are the $n$-scalar realisations found in [18], which correspond to the
renormalisations (3.10) of the classical currents (3.2) with $d_{ijk}$ given by
(3.6).  At $c=100$, these take the form
$$
T=T_X+ \ft12 (\partial\varphi_1)^2+\ft12(\partial\varphi_2)^2 +
\sqrt\hbar \big( \alpha_1\partial^2\varphi_1 +\alpha_2\partial^2\varphi_2
 \big)\eqno(3.14a)
$$
$$
\eqalignno{
W&={2\over \sqrt{261}}\Big\{  \ft13(\partial\varphi_1)^3
-\partial\varphi_1 (\partial\varphi_2)^2 +\sqrt\hbar \big(\alpha_1
\partial\varphi_1  \partial^2\varphi_1 -2\alpha_2 \partial\varphi_1
\partial^2\varphi_2 - \alpha_1 \partial\varphi_2 \partial^2 \varphi_2 \big)\cr
&\qquad\qquad +\hbar\big( \ft13\alpha_1^2\partial^3\varphi_1
-\alpha_1\alpha_2 \partial^3\varphi_2 \big)
-2\partial\varphi_1\, T_X -\alpha_1 \sqrt\hbar\,  \partial T_X\Big\},&(3.14b)
\cr}
$$
where $T_X$ is a stress tensor for $D=n-2$ scalar fields without background
charges,
$$
T=\ft12 \sum_{\mu=1}^D \partial X^\mu\, \partial X^\mu,\eqno(3.15)
$$
and the background charges $\alpha_1$ and $\alpha_2$ for $\varphi_1$ and
$\varphi_2$ are given by
$$
\eqalign{
\alpha_1^2&=-\ft{49}{8}\cr
\alpha_2^2&=\ft1{12}(D-\ft{49}2).\cr}\eqno(3.16)
$$
These conditions on the background charges ensure that the matter central
charge satisfies
$$
c=D+(1-12\alpha_1^2)+(1-12\alpha_2^2)=100.\eqno(3.17)
$$
Note that no matter how many scalar fields one chooses, including $n=100$, it
is necessary to have background charges in order to achieve $c=100$.

     Now that we have obtained a nilpotent BRST operator, and appropriate
matter realisations of the $W_3$ algebra, it is completely straightforward to
write down a Lagrangian for anomaly-free $W_3$ gravity.  We shall not give the
detailed result here; it may be found in [10].  Here, we just remark that it is
obtained from the general BRST prescription:
$$
\lagr=\ft12\bar\partial\varphi^i\partial\varphi^i
 -h T -B W
+\delta\Big( b(h-h_{\rm back}) +\beta(B-B_{\rm back})\Big),
\eqno(3.18)
$$
where $\delta$ denotes the BRST variation, which can be deduced from the BRST
operator (3.11), and $h_{\rm back}$ and $B_{\rm back}$ denote background
gauge-fixed values for the spin-2 and spin-3 gauge fields $h$ and $B$.  Thus
one has [10]
$$
\eqalign{
\lagr&=\ft12\bar\partial\varphi^i\partial\varphi^i
-b{\bar\partial c} -\beta {\bar
\partial}\gamma \cr &\quad +\pi_b(h-h_{\rm back}) + \pi_\beta(B-B_{\rm back})
- h(T+T_{\rm gh}) - B(W+W_{\rm gh}),\cr
}\eqno(3.19)
$$
where $\delta b=\pi_b$ and $\delta \beta=\pi_\beta$.  As in section 2, the
$\hbar$-independent terms in (3.19) (with matter currents $T$ and $W$ given by
(3.14$a$,$b$), and ghost currents $T_{\rm gh}$ and $W_{\rm gh}$ given by
(3.13$a$,$b$)) represent the classical Lagrangian (plus ghost Lagrangian), and
the $\hbar$-dependent terms correspond to counterterms necessary for the
explicit removal of the potential anomalies.

\bigskip
\noindent{\bf 4. Discussion}
\bigskip

     In this paper, we have reviewed some of the aspects of the quantisation of
$W_\infty$ and $W_3$ gravities.  Our discussion has been concerned entirely
with {\it chiral} $W$ gravity, in the sense that we have considered gaugings
only of a single (left-moving) copy of the algebra.  As remarked at the
beginning of section 2, it is completely straighforward to extend all of the
discussions in this paper to the non-chiral case by exploiting the ingenious
trick, introduced in [9], of using additional, auxiliary, fields.  Thus, for
example, for $W_3$ gravity we introduce auxiliary fields $J^i$ and $\tJ^i$, and
write the classical Lagrangian as
$$
\eqalign{
\lagr&=-\ft12\bar\partial\varphi^i\partial\varphi^i-J^i\tJ^i
+\tJ^i\partial\varphi^i+J^i\bar\partial\varphi^i\cr
&\quad -\ft12hJ^iJ^i-\ft13Bd_{ijk}J^iJ^jJ^k-\ft12\th\tJ^i\tJ^i-\ft13\tB
d_{ijk}\tJ^i\tJ^j\tJ^k,\cr}\eqno(4.1)
$$
where the tilded variables refer to a second (right-moving) copy of the gauge
algebra. The equations of motion for the auxiliary fields are
$$\eqalign{J^i&=\partial\varphi^i
-\th \tJ^i -\tB d_{ijk}\tJ^j \tJ^k,\cr
\tJ^i&=\bar\partial\varphi^i
-h J^i -B d_{ijk} J^j  J^k,\cr}\eqno(4.2)
$$
which can be recursively solved to give
$J^i$ and $\tJ^i$ as non-polynomial expressions in $\varphi^i$ and the gauge
fields.  Upon quantisation, one finds that the auxiliary fields have the
propagators
$$
\eqalign{
J^i(z)J^j(w)&\sim {\hbar\delta^{ij}\over (z-w)^2},\cr
\tJ^i(\bar z)\tJ^j(\bar w)&\sim {\hbar\delta^{ij}\over (\bar z-\bar w)^2},\cr
J^i(z)\tJ^j(\bar w)&\sim 0.\cr}\eqno(4.3)
$$
Thus the whole problem has been cloven into separate left-moving and
right-moving sectors.  The left-moving matter and ghost currents are now
constructed, at the full quantum level, by replacing $\partial\varphi^i$ in
(3.10), {\it etc}, by $J^i$.  Similarly, one uses $\tJ^i$ in the construction
of analogous right-moving currents.  Full details may be found in [11].

     An obvious application for anomaly-free $W_3$ gravity is in the
construction of the $W_3$ extension of string theory, {\it i.e.}\ $W_3$
strings. The idea is that the equations of motion for the spin-2 and spin-3
gauge fields impose the vanishing of the spin-2 and spin-3 currents.  At the
quantum level, these conditions can be interpreted, as in ordinary string
theory, as operator constraints on physical states.  By interpreting the
scalar fields in the matter realisations (3.14$a$,$b$) as spacetime
coordinates, one arrives at a first-quantised  description of $W_3$-string
excitations in an $n$-dimensional spacetime.  Because of the necessity for
background charges one does not have the full $SO(1,n-1)$ Lorentz group
acting, but only $SO(1,n-3)$.  The issues arising in the analysis of the
spectrum of $W_3$ strings are quite involved. Preliminary discussions were
given in [11], and a more extensive analysis is contained in [23].  Here, we
just summarise a couple of the main results.

     One can see from the form of the realisations (3.14$a$,$b$) that the
scalar $\varphi_1$ is on a very special footing.  In fact all the remaining
scalars ($\varphi_2$ and $X^\mu$) appear only {\it via} their stress tensor.
(In the case of $\varphi_2$, it has a background-charge contribution in the
stress tensor.)  Thus, in some sense the scalar $\varphi_1$ is the only one
which is intrinsically ``non-stringy'' in nature.  The spin-3 current $W$ can
be written as a sum of terms that involve only $\varphi_1$, plus terms
involving the total stress tensor $T$.  It is then rather easy to see at the
classical level that having imposed the $T=0$ constraint, the $W$ constraint
reduces to the statement that $\varphi_1={\rm constant}$ [11].  At the
first-quantised level, this becomes the statement that physical states cannot
involve any $\alpha_{-n}$ creation operators in the $\varphi_1$ direction [23].
The conclusion from this is that the sole effect of the $W$ constraint is to
``freeze out'' the $\varphi_1$ coordinate:  One is left with a theory that
looks remarkably like ordinary string theory.  The generalisation to $W_3$
strings  introduces the new feature of the non-stringy coordinate $\varphi_1$,
but this is then removed again by the new $W$ constraint, so it seems that the
Lord giveth,  and the Lord taketh away.  Unfortunately, it seems that one is
left with a string-like theory that lacks certain of the attractive features
of normal string theory; in particular, it seems not to describe any
massless states [23].  Of course one may hope that this aspect is not
generic to all higher-spin extensions.  In particular, it would be
interesting to see what happens for a supersymmetric extension of the $W_3$
algebra.  This is currently under investigation [24].

     There are other aspects of the quantisation of $W$-gravity theories that
we have not touched on here.  In particular, there is the very interesting
problem of constructing the $W_3$ analogue of the Polyakov induced action of
two-dimensional gravity [25].  To do this, one wants to choose a matter
realisation with non-critical value for the central charge, so that the
universal anomaly ``brings to life'' the Liouville field (and its higher-spin
analogues).  Considerable progress in this area has been made [26]. There seems
to be a certain sense in which there is really no such thing as a
``non-critical'' theory, since the Liouville fields will always come to the
rescue and make up the deficit in the central charge.  It would be interesting
to see whether there is ultimately a convergence of the critical and
non-critical approaches.  Similar issues have also been considered for
$W_\infty$ and $W_{1+\infty}$ gravity.  In particular, it has been shown
that the hidden $SL(2,R)$ Kac-Moody symmetry of light-cone two-dimensional
gravity [27] generalises to $SL(\infty,R)$ for the $W_\infty$ case, and
$GL(\infty,R)$ for the $W_{1+\infty}$ case [28].

     Two-dimensional gravity has proved to be an exceptionally interesting
field of study.  $W$ gravity promises to be an even richer one.

\np
\centerline{\bf Acknowledgments}

     I am very grateful to all my collaborators in the work described in this
review: Eric Bergshoeff, Paul Howe, Hong Lu, Larry Romans, Ergin Sezgin, Shawn
Shen, Kelly Stelle, Xujing Wang, Kaiwen Xu and Kajia Yuan.  I am also very
grateful to the Theory Group at Imperial College, London; the Theory Division
at CERN; and the ICTP, Trieste, for hospitality during the course of some of
this work, and to the organisers of the {\it Strings and Symmetries} meeting at
Stony Brook, May 1991.

\bigskip\bigskip

\singlespace
\centerline{\bf REFERENCES}
\frenchspacing
\bigskip

\item{[1]}A.B.\ Zamolodchikov, {\sl Teo.\ Mat.\ Fiz.}\ {\bf 65} (1985) 347.

\item{[2]}V.A.\ Fateev and A.\ Zamolodchikov, {\sl Nucl.\  Phys.}\  {\bf
B280} (1987) 644;\nl
V.A.\ Fateev and S.\ Lukyanov,  {\sl Int.\ J.\ Mod.\  Phys.}\ {\bf
A3} (1988) 507.

\item{[3]}C.N.\ Pope, L.J.\ Romans and X.\ Shen, {\sl Phys.\ Lett.}\ {\bf 236B}
(1990) 173;\nl
{\sl Nucl.\ Phys.}\ {\bf B339} (1990) 191.

\item{[4]}C.N.\ Pope, L.J.\ Romans and X.\ Shen, {\sl Phys.\ Lett.}\ {\bf
254B} (1991) 401.

\item{[5]}I.\ Bakas, {\sl Phys.\ Lett.}\ {\bf 228B} (1989) 57.

\item{[6]}E.\ Bergshoeff, C.N.\ Pope, L.J.\ Romans, E.\ Sezgin, X.\ Shen and
K.S.\ Stelle, {\sl Phys.\ Lett.}\ {\bf 243B} (1990) 350.

\item{[7]}E.\ Bergshoeff, P.S.\ Howe, C.N.\  Pope, E.\ Sezgin, X.\ Shen and
K.S.\ Stelle, {\sl Nucl.\ Phys.}\ {\bf B363} (1991) 163.

\item{[8]}C.M.\ Hull, {\sl Phys.\ Lett.}\ {\bf 240B} (1989) 110.

\item{[9]}K.\ Schoutens, A.\ Sevrin and P.\ van Nieuwenhuizen, {\sl Phys.\
Lett.}\ {\bf B243} (1990) 245.

\item{[10]}C.N.\ Pope, L.J.\ Romans and K.S.\ Stelle, ``Anomaly-free $W_3$
gravity and critical $W_3$ strings,'' preprint CERN-TH.6171/91, {\sl Phys.\
Lett.}\ {\bf B} (in press).

\item{[11]}C.N.\ Pope, L.J.\ Romans and K.S.\ Stelle, ``On $W_3$  strings,''
preprint CERN-TH.6171/91, {\sl Phys.\  Lett.}\ {\bf B} (in press).

\item{[12]}E.\ Bergshoeff, C.N.\ Pope, L.J.\ Romans, E.
 Sezgin and X.\ Shen, {\sl Phys.\ Lett.}\ {\bf 245B} (1990) 447.

\item{[13]}C.N.\ Pope, L.J.\ Romans and X.\ Shen, {\sl Phys.\ Lett.}\ {\bf
245B} (1990) 72.

\item{[14]}E.\ Bergshoeff, B.\ de Wit and M.\ Vasiliev, {\sl Phys.\ Lett.}\
{\bf
256B} (1991) 199;\nl
``The structure of the super-$W_\infty(\lambda)$ algebra,'' preprint,
CERN TH-6021-91.

\item{[15]}M.\ Fukuma, H.\ Kawai and R.\ Nakayama, ``Infinite-dimensional
Grassmannian structure of two-dimensional quantum gravity,'' preprint, UT-572.

\item{[16]}K.\ Yamagishi, {\sl Phys.\ Lett.}\ {\bf 266B} (1991) 370.

\item{[17]}C.N.\ Pope, L.J.\ Romans and X.\ Shen, {\sl Phys.\ Lett.}\ {\bf
254B} (1991) 401.

\item{[18]}L.J.\  Romans, {\sl Nucl.\  Phys.}\ {\bf B352} (1991) 829.

\item{[19]}F.\  Bais, P.\  Bouwknegt, M.\ Surridge and K.\  Schoutens, {\sl
Nucl.\ Phys.}\ {\bf B304} (1988) 348; 371.

\item{[20]}N.\ Mohammedi, ``On gauging and realising classical and quantum
$W_3$
symmetry,''  ICTP preprint, IC-91-49.

\item{[21]}J.\  Thierry-Mieg, {\sl Phys.\ Lett.}\  {\bf 197} (1987) 368.

\item{[22]}K.\ Schoutens, A.\ Sevrin and P.\  van Nieuwenhuizen, {\sl Comm.\
Math.\ Phys.}\ {\bf 124} (1989) 87.

\item{[23]}C.N.\ Pope, L.J.\ Romans, E.\ Sezgin and K.S.\ Stelle, ``The
$W_3$ string spectrum,'' preprint, CTP TAMU-68/91.

\item{[24]}H.\ Lu, C.N.\ Pope, L.J.\ Romans, X.J.\ Wang and K.W.\ Xu, in
progress.

\item{[25]}A.M.\ Polyakov, {\sl Mod.\ Phys.\ Lett.}\ {\bf A2} (1987) 893.

\item{[26]}H.\ Ooguri, K.\ Schoutens, A.\ Sevrin and P.\ van Nieuwenhuizen,
``The induced action of $W_3$ gravity,'' preprint, ITP-SB-91-16.

\item{[27]}V.G. Knizhnik, A.M.\ Polyakov and A.B.\ Zamolodchikov, {\sl
Mod.\ Phys.\ Lett.}\ {\bf A3} (1988) 819.

\item{[28]}C.N.\ Pope, X.\ Shen, K.W.\ Xu and K.\ Yuan, ``$SL(\infty,R)$
symmetry of quantum $W_\infty$ gravity,'' preprint, CTP TAMU-37/91, and
Imperial/TP/90-91/29.

\end